\documentstyle[prb,aps,epsf,preprint]{revtex}
\begin{document}
% for two column  activate the line below...                
%\twocolumn[\hsize\textwidth\columnwidth\hsize\csname@twocolumnfalse\endcsname
%LA-UR-95-2706 
\title
{Magnetism and Superconductivity in (RE)Ni$_2$B$_2$C: 
The Case of TmNi$_2$B$_2$C }
\author{M.L. Kuli\'c$^{a,b}$, 
A. I. Buzdin$^b$, and L. N. Bulaevskii$^c$ }
\address{$^{a}$Max-Planck-Institut f\"{u}r Festk\"{o}rperforschung,
Heisenbergstr. 1,
70569 Stuttgart, Germany\\
$^{b}$Centre de Physique Th\'{e}orique et de Mod\'{e}lisation,\\
Universit\'{e} BordeauxI, CNRS-URA 1537 Gradignan Cedex, France \\
$^{c}$Los Alamos National Laboratory, Los Alamos, NM87545}
\date{\today}
\maketitle
\begin{abstract}
The recently reported \cite{Chang,Lyn} coexistence of an oscillatory
magnetic order with the wave vector $Q=0.241$ \AA$^{-1}$ and superconductivity
in TmNi$_{2}$B$_{2}$C is analyzed
theoretically. It is shown that the oscillatory magnetic order and
superconductivity interact predominantly via the exchange
interaction between localized moments (LM's) and conduction
electrons, while the electromagnetic interaction between them is negligible.
In the coexistence phase of the clean TmNi$_{2}$B$_{2}$C
the quasiparticle spectrum should have a line of zeros at the Fermi surface, giving
rise to the power law behavior of thermodynamic and transport properties. Two
scenarios of the origin of the oscillatory magnetic order in TmNi$_{2}$B$_{2}$C
are analyzed: a) due to superconductivity and b) independently on superconductivity. 
Experiments in magnetic field are proposed in order to
choose between them. \\
e-mail: kulic@audrey.mpi-stuttgart.mpg.de
\end{abstract}

% for two column  activate the line below... 
%]

\narrowtext

The problem of the coexistence of magnetic order and superconductivity is a
long-standing one and Ginzburg\cite{Ginzburg} was the first to note the
antagonistic character of these phenomenon. Further impetus in this field
came after the discovery of the ternary rare earth (RE) compounds 
(RE)Rh$_{4}$B$_{4}$ and (RE)Mo$_{6}$X$_{8}$ (X=S,Se), see Ref.~\onlinecite{IshiMap}. In many of
these compounds both ferromagnetic (F) and antiferromagnetic (AF)
ordering coexist with superconductivity (S), see Refs.~\onlinecite{BuBuKuPa,Maple}.
It turned out that the coexistence of S and AF ordering was realized in
many of these compounds\cite{Buzdin} usually up to $T=0$, while S and
modified F ordering coexisted in ErRh$_{4}$B$_{4}$, HoMo$_{6}$S$_{8}$ and 
HoMo$_{6}$Se$_{8}$ only. The reason for this was the antagonistic characters of
these orderings. A theory has been developed and the phase diagram was given in 
Refs.~\onlinecite{BuBuKuPa,BuBuKuPa1,BuRuKu} where the possibility of the
coexistence of S and spiral or domain-like magnetic order has been
elaborated quantitatively by including the exchange (EX) and
electromagnetic (EM) interaction of conduction electrons and localized
magnetic moments (LM's). It has been also demonstrated that the
theory based on the EM interaction only \cite{BloVar} can not describe the
coexistence problem in real systems. Note, some heavy fermions 
UPt$_{3}$,URu$_{2}$Si$_2$ etc. show a coexistence of the AF and S orderings. Recently, it
has been also found experimentally the coexistence of nuclear magnetism and
superconductivity\cite{Pobel} in AuIn$_{2}$, which was theoretically
analyzed in Ref.~\onlinecite{Kulic}.

However, recent discovery of superconductivity in the quaternary
intermetallic compounds (RE)Ni$_2$B$_2$C (RE=Sc,Y,Lu,Tm,Er,Ho and Th) has
received appreciable attention, because of the relatively high
superconducting transition temperature - $16.6$ K in Lu. Band structure
calculations \cite{Mattheiss} show that the electronic spectrum is three-dimensional.
Because of the spatial isolation of magnetic ions there is a possibility for
the coexistence of a magnetic (M) order and S in (Ho,Er,Dy)Ni$_2$B$_2$C 
with ($T_c;T_M$)=(8,11,6.5 K;6,7,10.5 K) respectively \cite{Zare}. $T_M$
is the (antiferro)magnetic transition temperature. These compounds are
characterized by the ferromagnetic alignment in each layer, with the
magnetic moments of two consecutive layers aligned in opposite directions.
Band structure calculations \cite{Rhee} of the nonmagnetic LuNi$_2$B$_2$C
compound show that the conduction electron density on the RE ions is small
(similarly as in (RE)Rh$_4$B$_4$ and (RE)Mo$_6$X$_8$ ) giving rise to a relatively
small exchange energy.

The subject of this paper is the theoretical analysis of the coexistence
problem in TmNi$_{2}$B$_{2}$C, which is superconducting below $T_c\approx 11$ K
with an oscillatory magnetic ordering of the Tm moments below $T_M\approx
1.5$ K with persisting coexistence up to $T=0$. The magnetic structure
is incommensurate \cite{Chang} with the Tm moments along the $c$-axis and
with a sinusoidal modulation of their magnitudes along the (110)
direction. This sinusoidal order is characterized by the wave vector $%
Q=0.241 $ \AA$^{-1}$. One should stress the following facts: (1) TmNi$_{2}$B$_{2}$C
is unique in the (RE)Ni$_2$B$_2$C family, which shows a modulation of the
magnetic order in the (110) direction and the alignment of the moments
along the $c$-axis; ($2$) the wave vector $Q$ is neither large nor small ($%
\xi _0^{-1},\lambda ^{-1}\ll Q\ll k_F$), where $\xi _0,\lambda ,k_F$ are the
superconducting coherence length, magnetic penetration depth and Fermi
momentum respectively.

In what follows the theory of magnetic superconductors - the MS theory 
\cite{BuBuKuPa,BuBuKuPa1}, is applied to TmNi$_{2}$B$_{2}$C and it will
be shown that the competition between S and the oscillatory M order is 
{\it predominantly} due to the EX interaction, while the EM one is
negligible, not only in this compound but also in the whole (RE)Ni$_{2}$B$_{2}$C
family. An analysis of effects of the magnetic field allows us to discern
between two possible scenarios for the origin of the oscillatory magnetic
order.

The MS theory \cite{BuBuKuPa,BuBuKuPa1} considers all important
interactions between $LM^{\prime }s$ and conduction electrons: (1) via the
direct EX interaction; (2) via the induced magnetic field ${\bf B}({\bf r}%
)={\rm curl}{\bf A}({\bf r})$ - the EM interaction, which is due to the dipolar
magnetic field ${\bf B}_m({\bf r})=4\pi {\bf M}({\bf r})$. The general
Hamiltonian of the (RE)Ni$_2$B$_2$C compounds is given by 
\[
\hat H=\int d^3r\{\psi ^{\dagger }({\bf r})\epsilon ({\bf \hat p-}\frac ec%
{\bf A})\psi ({\bf r})+[\Delta ({\bf r})\psi ^{\dagger }({\bf r})i\sigma
_y\psi ^{\dagger }({\bf r})+c.c]+\frac{\mid \Delta ({\bf r})\mid ^2}V+\hat H%
_{imp} 
\]
\begin{equation}  \label{eq.1}
+\sum_iI({\bf r}-{\bf r}_i)\psi ^{\dagger }({\bf r}){\bf \sigma }(g-1){\bf J}%
_i\psi ({\bf r})+\frac{(rot{\bf A}({\bf r}))^2}{8\pi }\}+\sum_i[-{\bf B}(%
{\bf r}_i)g_e\mu _B{\bf J}_i+\hat H_{CF}({\bf J}_i)].
\end{equation}
Here, $\epsilon ({\bf \hat p-}\frac ec{\bf A}),$ $\Delta ({\bf r}),$ ${\bf A,%
}$ $I({\bf r}),$ $V,$ ${\bf \sigma },$ ${\bf J}_i$ and $g$ are the
quasiparticle energy, the superconducting order parameter, the vector
potential, the exchange integral, the electron-phonon coupling constant,
Pauli matrices, the total angular moment and the Lande factor respectively.
The first three terms in Eq.~(1) describe the superconducting mean-field
Hamiltonian in the magnetic field ${\bf B}({\bf r})={\rm curl}{\bf A}({\bf r})$,
while the term $\hat H_{imp}$ describes the electron scattering on
nonmagnetic impurities. The term proportional to ${\bf \sigma }(g-1){\bf J}%
_i $ describes the direct EX interaction between electrons and $LM^{\prime
}s$, while $\hat H_{CF}({\bf J}_i)$ is responsible for the crystal field
effects and magnetic anisotropy. Based on Eq.~(1) and by using Eilenberger
equations one can find the free-energy functional of the coexistence phase in 
terms of order parameters -- see below.

A. Characteristic parameters of TmNi$_{2}$B$_{2}$C

The critical temperature of the transition into the oscillatory magnetic
state $T_{M}\approx 1.5$ K is small compared to the superconducting
critical temperature $T_{c}\approx 11.5$  K and it is of the order of the
exchange energy $\Theta _{ex}$ - see below. From $\Theta _{ex}=N(0)h_{ex}^{2}
$ we can estimate the exchange interaction between electrons and LM's, 
which is characterized by $h_{ex}=I(0)(g-1)nJ$ , where $N(0)$ is the
electronic density of states at the Fermi level (per LM), $n$ is the
concentration of LM's. In absence of data on $N(0)$ and Fermi velocity $%
v_{F}$ in TmNi$_{2}$B$_{2}$C we use the band structure value\cite{Mattheiss}
for LuNi$_{2}$B$_{2}$C, where $N(0)\simeq 2.4$ states/eV$\cdot$ Lu atom and $%
v_{Fx}=v_{Fy}\simeq (2-3)\times 10^{7}$ cm/sec. This procedure is
justified because the $f$-levels of Lu and Tm ions are weakly coupled to
the conduction electrons. The decrease of $T_{c}$ in the (RE)Ni$_{2}$B$_{2}$C 
family is scaled\cite{Eisaki} by de Gennes factor $(g-1)^{2}J^{2}$ which
allows to estimate $\Theta _{ex}=N(0)h_{ex}^{2}$ and $h_{ex}$. Namely, the
Abrikosov-Gorkov formula $dT_{c}/dx\simeq -\pi ^{2}\bar{\Theta}_{ex}/2$ for
the decrease of $T_{c}$ in Lu$_{1-x}$Tm$_{x}$Ni$_{2}$B$_{2}$C , i.e. $(\Delta
T_{c}/\Delta x)_{{\rm Lu-T_{m}}}\approx T_{c}^{{\rm Lu}}-T_{c}^{{\rm Tm}}\approx 5$ K$\sim 5%
\bar{\Theta}_{ex}$ gives $\bar{\Theta}_{ex}\sim 1$  K and $h_{ex}\sim 60$ K.

The long-range part of the EM dipole-dipole interaction between LM's 
is characterized by $\Theta _{em}=2\pi n\mu ^2$ , $\mu =g\mu
_BJ.$ The neutron diffraction measurements \cite{Chang} in TmNi$_{2}$B$_{2}$C give 
$\mu \simeq 5\mu _B,$ while from the crystallographic structure follows $%
n\approx 2\cdot 10^{22}$ cm$^{-3}$, which gives $\Theta _{em}\approx 2$  K.
Note that $\Theta _{em}$ $\sim \Theta _{ex}\sim T_M$. From $\Delta _0\simeq
1.76$ $T_c$ one obtains $\xi _0\simeq 250$ \AA, while from magnetization
measurements near $H_{c2}$ and from the slope of $H_{c2}$ near $T_c$ (see Ref.~\onlinecite
{Cho}) it follows $\kappa=\lambda /\bar \xi \simeq 7$ and $\bar \xi \simeq
110$ $A$, where $\lambda \approx 0.62\lambda _L(\xi _0/l)^{1/2}$ and $\bar %
\xi \approx 0.85(\xi _0l)^{1/2}$. This gives the mean-free path $l\approx 50$
\AA and the London penetration depth $\lambda _L\approx 500$ \AA. One can
say that the samples studied by Cho {\it et al.} \cite{Cho} were in dirty limit where also holds: $%
(h_{ex}\tau /\hbar )^2\ll 1$ and $(Ql)^{-2}\ll 1$. .

B. Free-energy functional of the coexistence phase: 

Since $Q\ll k_F$ the problem of interplay between S and M is treated\cite{BuBuKuPa,BuBuKuPa1} using 
the Eilenberger equations for the normal $g_\omega ({\bf v},{\bf R})$ and
anomalous $f_\omega ({\bf v},{\bf R})$ electronic Green's function. They 
describe the motion of electrons in the EX field $\vec h_{ex}({\bf R}%
)={\bf c}h_Q\sin Qz$ ($h_Q=h_{ex}S_Q$, $S_Q=|\langle{\bf J}\rangle|/J)$ -
the EX interaction, and in the dipolar magnetic field ${\bf B}({\bf r})={\rm curl}%
{\bf A}({\bf r})$ - the electromagnetic EM interaction. We present only
some necessary results for the free-energy (per LM) $F\{\Delta ,S_Q,{\bf Q}%
\}=F_s\{\Delta \}+F_M\{S_Q\}+F_{int}\{\Delta ,S_Q\}$, where 
\begin{equation}  \label{eq.2}
F_s\{\Delta \}=-\frac 12N(0)\Delta ^2\ln \frac{e\Delta _0^2}{\Delta ^2}.
\end{equation}
$\Delta $ is the $S$ order parameter and $\Delta _0$ is the $S$ order
parameter in equilibrium and in absence of magnetism. The magnetic part $F_M$ in the mean-field approach 
is given by 
\[
F_M\{S_Q\}=-\sum_Q\{[\Theta _0+\Theta _{ex}(\tilde \chi _e(Q)-1)][\mid {\bf S%
}_{Q,\perp }\mid ^2+\mid {\bf S}_{Q,\parallel }\mid ^2] 
\]
\begin{equation}  \label{eq.3}
+\Theta _{em}\mid {\bf S}_{Q,\parallel }\mid ^2+D(\mid S_{x,Q}\mid ^2+\mid
S_{y,Q}\mid ^2)\}+F_0\{{\bf S}_Q\}
\end{equation}
Here, $D>0$ and ${\bf S}_{Q,\perp }$, ${\bf S}_{Q,\mid \mid }$ are
transverse and longitudinal (w.r.t. ${\bf Q}$) components of ${\bf S}_Q$
respectively, while $F_0\{S_Q\}$ is the isotropic part (entropy term) of the
functional for isolated ions. $\Theta _0=\Theta _{ex}+\Theta _{em} /3+
\Theta' _{ex}+\Theta' _{em}$ characterizes the contribution of
all mechanisms (long ($\Theta _{ex},\Theta_{em}$)
- and short ($\Theta' _{ex},\Theta' _{em}$)-range parts of the exchange and dipole
energies) to the ground-state energy\cite{BuBuKuPa,BuBuKuPa1}. We assume that in the 
normal state the dipole-dipole interaction of LM moments leads to ferromagnetic ordering, but 
exchange interaction may result in ferromagnetic or oscillatory ordering depending on Fermi surface structure. The
electronic susceptibility $\chi _e(Q)\equiv N(0)\tilde \chi _e(Q)$ is still
unknown, although it can be calculated by knowing the band structure of TmNi$_{2}$B$_{2}$C. 
Two scenarios for $\chi _e(Q)$ will be presented below. The
interaction part, $F_{int}\{\Delta ,S_Q\}=F_{int}^{ex}+F_{int}^{em}$, of the
free-energy contains the EX and EM contributions respectively.

For samples TmNi$_{2}$B$_{2}$C studied by Cho {\it et al.} \cite{Cho} the {\it dirty limit} ($%
Q\ll l\ll \xi _{0}$) is realized, as well as $%
lh_{ex}S_{Q}/v_{F})^{2},(elA_{Q}/c)^{2}\ll 1$, which allows us to find $%
F_{int}$ 
\begin{equation}
F_{int}\{\Delta ,S_{Q}\}\equiv \frac{\pi N(0)\Delta }{2\tau _{m}}=\frac{\pi
^{2}\Delta }{2}\sum_{Q}\{\frac{\Theta _{ex}}{v_{F}Q}\mid {\bf S}_{Q}\mid
^{2}+\frac{3\Theta _{m}}{2v_{F}\lambda _{L}^{2}Q^{3}}\mid {\bf S}_{Q,\perp
}\mid ^{2}\}.  \label{eq.4}
\end{equation}
The first term on the right hand side of Eq.~(4) describes $F_{int}^{ex}$ and the
second one $F_{int}^{em}$ respectively. Eq.~(5) is derived by assuming
that: (a) $\tau _{m}\Delta >1$ what is indeed fulfilled in TmNi$_{2}$B$_{2}$C,
where $\tau _{m}\Delta \approx 6-7$; (b) the Fermi surface is isotropic -
fulfilled also in TmNi$_{2}$B$_{2}$C. Note, the expression for $%
F_{int}\{\Delta ,S_{Q}\}$ in the clean limit can be found in Ref.~\onlinecite{BuBuKuPa}.

Already on this level we can estimate the relative contribution of the EX
and EM terms in the interaction of superconducting and magnetic subsystems
in TmNi$_{2}$B$_{2}$C. For parameters extracted from experiments - see A,
one gets $r\equiv (F_{int}^{em}/F_{int}^{ex})\simeq \Theta _{em}/\Theta
_{ex}(\lambda _LQ)^2\approx 10^{-4}$. This important result means that in TmNi$_{2}$B$_{2}$C 
the EM interaction makes a {\it negligible} contribution
to $F_{int}$ and as a consequence the competition of S and the M order in TmNi$_{2}$B$_{2}$C 
is exclusively due to the {\it EX interaction} . The similar
situation is realized in the whole family of (RE)Ni$_2$B$_2$C compounds where $%
r<10^{-3}$. Moreover, when TmNi$_{2}$B$_{2}$C is placed in external magnetic field
the EX interaction plays a decisive role- see below. This means that the
approach which is based on the EM interaction only (see Ref.~\onlinecite{Ng} and
references therein) is inadequate in explaining properties of the 
(RE)Ni$_2$B$_2$C family. However, the EM 
interaction, although much less detrimental for superconductivity than the EX 
one, makes the magnetic structure transverse, i.e. ${\bf S\cdot Q=0}$
due to the $\Theta _{em}\mid {\bf S}_{Q,\parallel }\mid ^2$ term in Eq.~(3).

C. Origin of the oscillatory magnetic order

The magnetic free-energy in the normal state $F_M\{S_Q\}$ depends on the
electronic susceptibility $\tilde \chi _e(Q)=\chi _e(Q)/g_e^2\mu _B^2N(0)$
and contains magnetic anisotropy ($D>0$) and the single ion term $F_0\{S_Q\}$
- see Eq.~(3). These quantities determine magnetic structure in TmNi$_{2}$B$_{2}$C 
in absence of superconductivity. At present both are unknown and therefore
in what follows we analyze two possible scenarios for the origin of the
oscillatory magnetic structure, which depends on the form of $\chi _e(Q)$ in 
TmNi$_{2}$B$_{2}$C:

\underline{$1.$ Ferromagnetic (F) scenario} In this scenario $\chi _{e}(Q)$
reaches maximum at $Q=0$ , i.e. the {\it ferromagnetic} order would be
realized in the normal conduction state of TmNi$_{2}$B$_{2}$C below some
temperature - see Fig.~1a. However, in the S state it is transformed into
an oscillating magnetic structure with the wave vector $Q\ll k_{F}$ - see
below, and because $a^{2}Q^{2}\ll 1$ (magnetic length $a<k_{F}^{-1}$ ) one has $\chi
_{e}(Q)\approx N(0)(1-a^{2}Q^{2})$. Replacing this $\chi _{e}(Q)$ 
in Eq.(3) and by minimizing the free-energy $F$ with respect to $Q$
one gets the sinusoidal magnetic structure at $T$ very neat $T_{M}$ (when $%
h_{ex}S_{Q}\ll \Delta $ and higher order terms in $\mid {\bf S}_{Q}\mid ^{2}$
are negligible ) with $Q_{M}=(\pi \Theta _{ex}/4\Theta _{0}a^{2}\xi
_{0})^{1/3}$. In the presence of the magnetic anisotropy and by lowering
temperature $\mid {\bf S}_{Q}\mid $ grows and higher order terms in $\mid 
{\bf S}_{Q}\mid ^{2}$ (described by $F_{0}\{{\bf S}_{Q}^{2}\}$ in Eq.~(3))
become important giving rise to higher harmonics $3Q,5Q,$ etc.. As a result
the striped transverse one-dimensional domain structure (${\bf S}_{Q}\cdot 
{\bf Q}=0$, ${\bf S}_{Q}\parallel z$-axis) is formed\cite{BuBuKuPa},\cite
{BuBuKuPa1} with the magnetic energy (per LM) 
\begin{equation}
F_{M}=F_{0}\{S_{Q}^{2}\}-\Theta _{ex}S_{Q}^{2}+\eta (S_{Q}^{2},T)\frac{Q}{%
\pi },  \label{eq.5}
\end{equation}
where $\eta $ is the domain wall energy given by $\eta =k_{F}^{-1}\Theta
_{w}S_{Q}^{2}$. Here, $\Theta _{w}\approx 0.6(\Theta _{0}D)^{1/2}$ for $%
D<\Theta _{0}$ but $(D/\Theta _{0})^{3/4}>0.25(k_{F}\xi _{0})^{-1/2}$, while 
$\Theta _{w}\approx 0.3\Theta _{0}$ for $D>\Theta $. This phase is in
further called the DS-phase. In the DS-phase and at $T\ll T_{M}$ the
wave vector of the structure is given by $Q_{DS}\approx 2(\Theta _{ex}/\Theta
_{w}k_{F}\xi _{0})^{1/2}$. Since $Q^{\exp }=0.241$ \AA$^{-1}$ and by knowing $%
k_{F}$ - for instance $k_{F}\sim 1$ \AA$^{-1}$, one obtains reasonable value
for $\Theta _{w}\sim (0.1-0.2)$  K. These results mean that in the F-scenario
the transformation from sinusoidal to the domain-like structure takes place
around $T_{M}$, with small changes from $Q_{M}$ to $Q_{DS}$, where
superconductivity and the domain-like magnetic structure (the DS phase)
coexist. Moreover, for the given set of parameters in TmNi$_{2}$B$_{2}$C - see 
A, one gets that at $T=0$ one has $F_{DS}-F_{M}\approx
-0.3N(0)\Delta ^{2}/2$, where $F_{DS}$ is the free-energy of the DS-phase.
This means that the $F$-scenario for the origin of the oscillatory magnetic
order in TmNi$_{2}$B$_{2}$C predicts that superconductivity {\it coexists%
} with the domain-like magnetic order up to $T=0$. We pay attention
that the latter result is independent of the scenario (F or O - see below)
and it is in accordance with the experimental finding\cite{Chang} in TmNi$_{2}$B$_{2}$C, 
where $S$ and the oscillatory magnetic order coexist up to $%
T=0$.

$II.$ \underline{O-scenario} - In this scenario it is assumed that the
{\it oscillatory} magnetic order with the wave vector $Q$ would be realized in
absence of superconductivity, i.e. $\chi _{e}(Q)$ is peaked at $Q_{0}$ - see 
Fig.~1b. At lower temperatures the magnetic anisotropy and the single ion
term $F_{0}\{S_{Q}^{2}\}$ transform the structure into a domain-like one. In
this case $F_{int}\{\Delta ,{\bf S}_{Q},{\bf Q}\}$ is also given by Eq.~(4),
where the wave-vector $Q$ should be considered fixed (by experiment). We
point out that the domain-like magnetic structure in the O-scenario is a
property of the normal state and not of the superconducting one - see $Fig.1b
$. Note, the ratio $r\ll 1$, i.e. it is small in both scenarios.
Because in the O-scenario one has also $F_{DS}-F_{M}\approx -0.3N(0)\Delta
^{2}/2$ then the domain-like magnetic structure and superconductivity
coexist also up to $T=0$.
\\[2.8cm]
%\begin{figure}[tbp]
\epsfysize=2.8in
\hspace*{0.0cm}
\epsffile{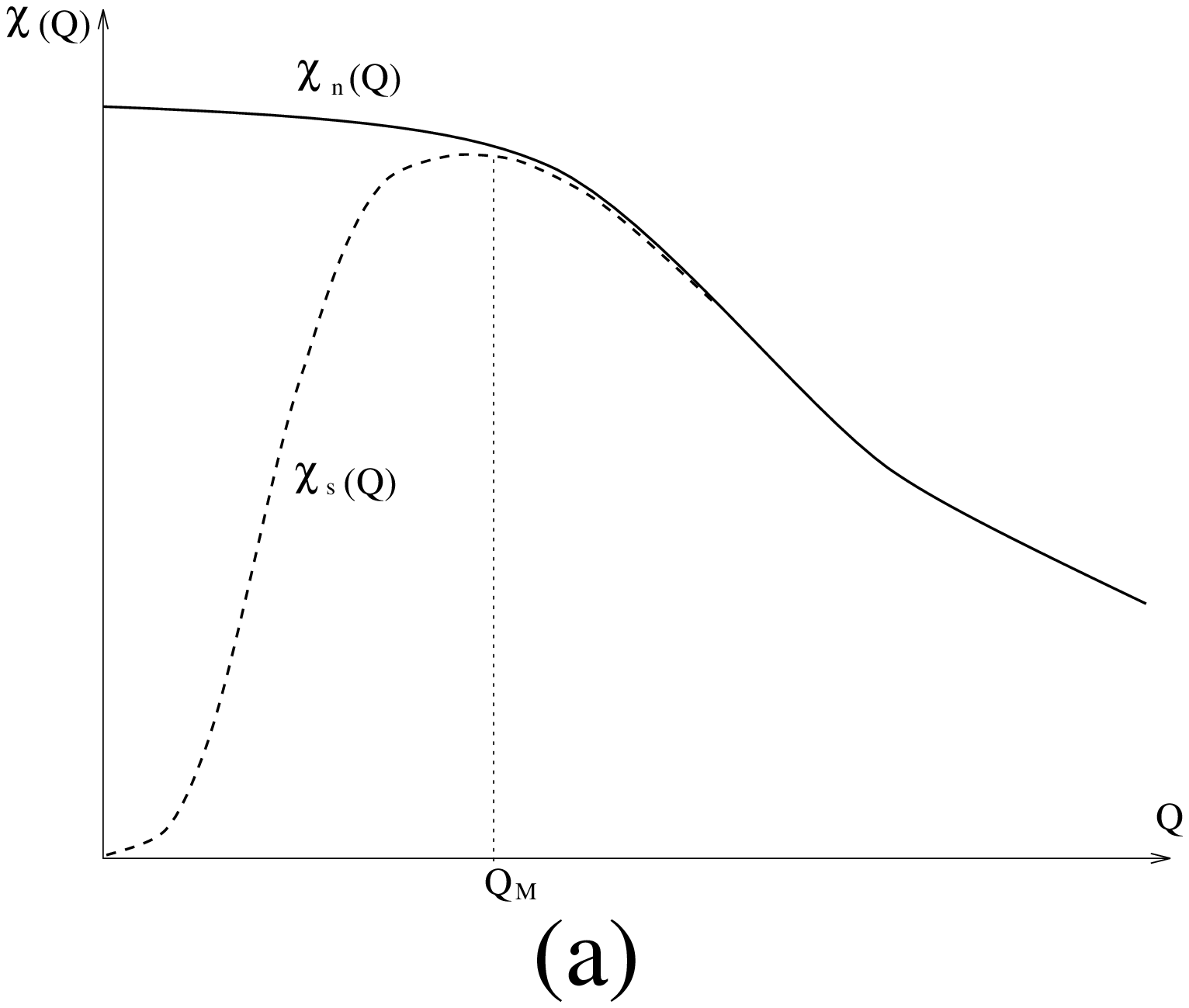}\\[-7.1cm]
\epsfysize=2.8in
\hspace*{8.6cm}
\epsffile{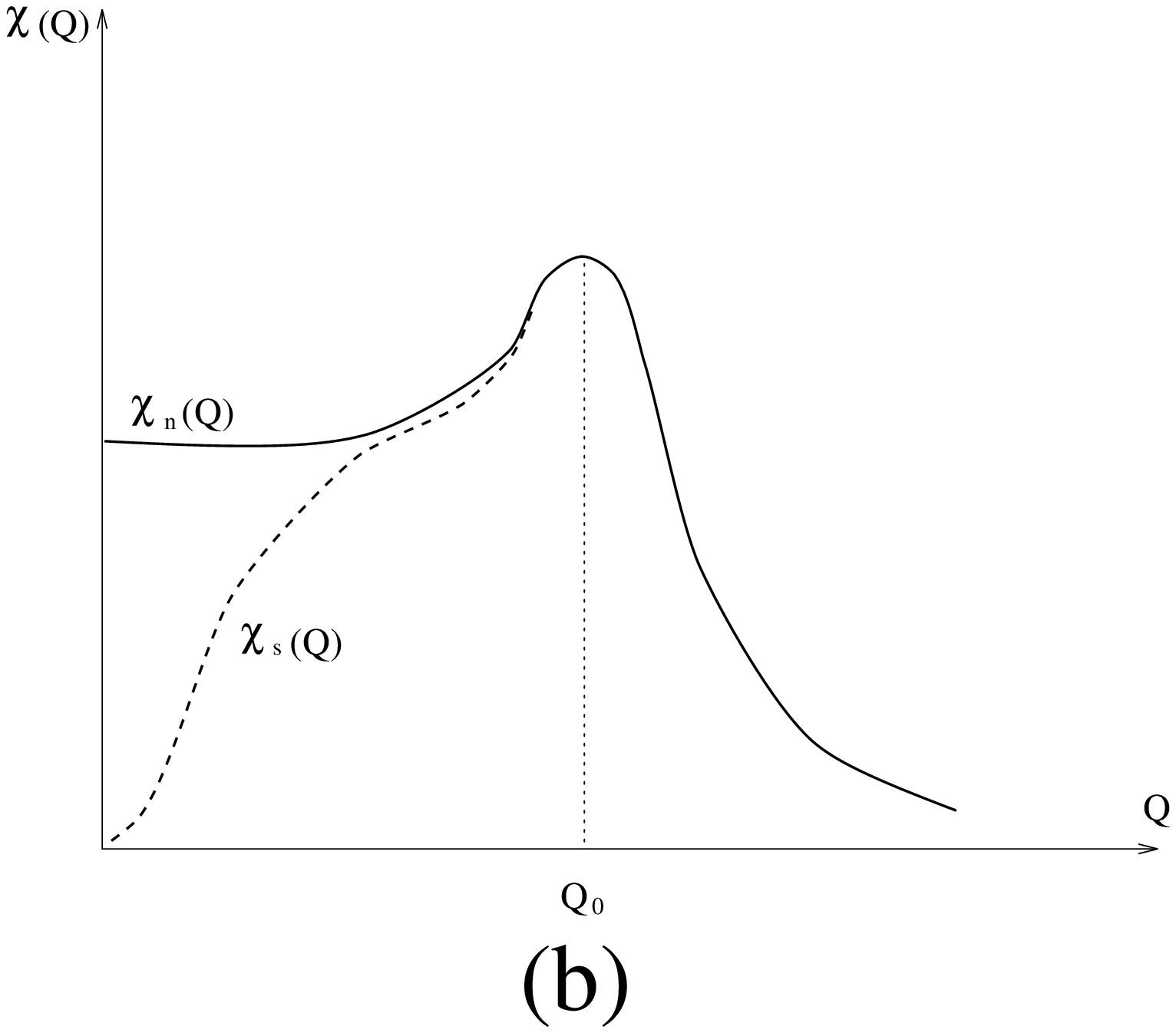}\\[-3cm]
%\caption{
FIG.~1. Schematic shape of the spin susceptibility of conduction electrons in
the normal - $\chi_n(Q)$ and superconducting -
$\chi_s(Q)$ state in the case of: (a) the F-scenario, where the
peak in $\chi_s(Q)$ is due to superconductivity; (b) the O-scenario, where 
the peak in $\chi_s(Q)$ is independent on superconductivity.
%} \end{figure}

Note, recent neutron scattering measurements show\cite{Lyn} clearly the
presence of the third harmonic ($3Q$), at $T<1$  K, with the intensity $%
I_{3Q}\approx 0.03I_Q$ what tells us that the magnetic structure in TmNi$_{2}$B$_{2}$C 
is domain-like. The smallness of $I_{3Q}$ can be due to the
presence of defects in the sample which are always detrimental for a domain
structure \cite{BuBuKuPa,BuBuKuPa2}. The F- and O-scenario can
not be resolved by this type of measurement. For that we need to study the
system in magnetic field.

\newpage

D. Gapless superconductivity

In clean superconductors the oscillatory magnetic order can give rise to the
gapless quasiparticle spectrum \cite{BuBuKuPa,BuBuKuPa1,BuRuKu}
if $h_{ex}>\Delta $, what is just the case in TmNi$_{2}$B$_{2}$C where $%
h_{ex}>60$  K and $\Delta <20$. The gapless region on the Fermi surface
given by the condition ${\bf Q\cdot v}_{F}=0$. In the $DS$-phase at
temperatures where $h_{ex}S_{Q}>\Delta $ the density of states is given by 
\begin{equation}
N_{s}(E)=N(0)\frac{h_{ex}S_{Q}}{\Delta \cdot v_{F}Q}E \ln \frac{4\Delta 
}{\pi E},\ \ \ \ E\ll \Delta .  \label{eq.6}
\end{equation}

By measuring tunneling conductance, where $\sigma (V)$ $\sim N_s(V)$, one
could test this prediction which is a consequence of the EX interaction.

E. Effects in magnetic field

Measurements in magnetic field can discern between $F$- and $DS$-scenarios.
In the $F$-scenario the critical field $H_{c}^{FS}$ for the first order $F$-$%
DS$ transition (at $T\ll T_{M}$) is obtained from the condition $%
-HM(0)=-N(0)\Delta _{0}^{2}/2$, i.e. $H_{c}^{FS}\approx 200$ G for $%
M(0)=10^{3}$ G ($M(0)=n\mu $ is the saturation magnetization). Note, in
getting $H_{c}^{FS}$ we have assumed that the field is oriented along the $c$%
-easy axis while if it is along the hard axis it could be much higher, i.e. $%
H_{a-b}^{FS}\gg H_{c}^{FS}.$ On the first sight such a small value of $%
H_{c}^{FS}$contradicts reports on the critical field\cite{Eisaki} in TmNi$_{2}$B$_{2}$C, 
where rather high critical field $H_{c2}\sim 1$ T is
found near $T_{M}$. Concerning this point one should stress the measurements
 \cite{Eisaki} have been done: (1) on the polycrystals; (2) at fixed
magnetic field $H$ by lowering temperature. Because of possibility $%
H_{a-b}^{FS}\gg H_{c}^{FS}$ the measurements on single crystals are desired,
and because the transition at $H_{c}^{FS}$ is of the first order, with a
possibility for huge hysteresis, one should perform measurements at fixed $T$
in increasing and decreasing field. Note, such a huge hysteresis is not
expected in the O-scenario, where the critical magnetic field is determined by superconducting 
properties mainly and must be
much larger than $H_{c}^{FS}$. 

In conclusion, we have found that the oscillatory (domain-like) magnetic
order and superconductivity coexist in TmNi$_{2}$B$_{2}$C up to $T=0$ and that
their competition is due to the exchange interaction between conduction
electrons and localized moments (LM's), while the electromagnetic
interaction is negligible, eventually helping in making the magnetic
structure transverse. The type and the origin of the magnetic structure in
absence of superconductivity can be resolved by measuring critical magnetic
fields, where a huge hysteresis could favor the $F$-scenario, i.e. the
ferromagnetic order in absence of superconductivity. It is predicted the
gapless superconductivity in clean TmNi$_{2}$B$_{2}$C with line of zeros on the
Fermi surface.

One of the authors (M.L.K.) would like to thank
Universit\'e Bordeaux for kind hospitality and O. Andersen, L. Hedin, Y.
Leroyer, M. Mehring and V. Oudovenko for support. Work in Los Alamos is 
supported by the U.S. DOE.

\end{document}